\documentclass[10pt]{article}

\usepackage{amsmath}
\usepackage{amssymb}
\usepackage{graphics}

\title{\textbf{Why Quantum Theory?}}

\author{Lucien Hardy\thanks{\texttt{hardy@qubit.org.}}\\
\textit{Centre for Quantum Computation,}\\
\textit{The Clarendon Laboratory,}\\
\textit{Parks road, Oxford OX1 3PU, UK}}

\begin{document}

\setlength{\textwidth}{12.5cm}
\setlength{\textheight}{19cm}

\maketitle

\begin{abstract}
The usual formulation of quantum theory is rather abstract.  In recent
work I have shown that we can, nevertheless, obtain quantum theory from
five reasonable axioms.  Four of these axioms are obviously consistent
with both classical probability theory and quantum theory.  The
remaining axiom requires that there exists a continuous reversible
transformation between any two pure states.  The requirement of
continuity rules out classical probability theory.  In this paper I will
summarize the main points of this new approach.  I will leave out the
details of the proof that these axioms are equivalent to the usual
formulation of quantum theory (for these see reference \cite{Hardy1}).
\end{abstract}

\section{Introduction}

The usual formulation of quantum theory is
very obscure employing complex Hilbert spaces, Hermitean operators and
so on.  While many of us, as professional quantum theorists, have
become very familiar with the theory, we should not mistake this familiarity
for a sense that the formulation is physically  reasonable.
Quantum theory, when stripped of all
its incidental structure, is simply a new type of probability theory.
Its predecessor, classical probability theory, is very intuitive.  It
can be developed almost by pure thought alone employing only some very
basic intuitions about the nature of the physical world.  This prompts
the question of whether quantum theory could have been developed in a
similar way. Put another way, could a nineteenth century physicist have
developed quantum theory without any particular reference to
experimental data?  In a recent paper
I have shown that the basic structure of quantum theory
for finite and countably infinite dimensional Hilbert spaces follows
from a set of five reasonable axioms \cite{Hardy1}.  Four of these
axioms are obviously consistent with both classical probability theory
and with quantum theory.  The remaining axiom states that there exists a
continuous reversible transformation between any two pure states.  This
axiom rules out classical probability theory and gives us quantum
theory.  The key word in this axiom is the word ``continuous''.  If it
is dropped then we get classical probability theory instead.
The proof that quantum theory follows from these axioms, although
involving simple mathematics, is rather lengthy.  In this paper I will
simply discuss the main ideas referring interested readers
to the main paper \cite{Hardy1}.

Various authors have set up axiomatic formulations of quantum
theory, for example see references
\cite{birkoff,mackey,piron,ludwig,mielnik,lande,fivel,accardi,landsman,cmw}
(see
also \cite{gleason,kochen,pitowsky}). Much of this work is in the quantum logic
tradition.  The advantage of the present work is that there
are a small number of simple axioms which can be easily
motivated without any particular appeal to experiment,
and, furthermore, the mathematical methods required to obtain quantum
theory from
these axioms are very straightforward (essentially just linear algebra).

\section{Basic notions}\label{basicnotions}

We will consider
situations in which a preparation apparatus prepares systems which may
be transformed by a transformation apparatus and measured by a
measurement apparatus.  Associated with any given preparation will be a
{\it state}.  {\it The state is defined to be (that thing described by) any
mathematical object that can be used to determine the probability
associated with each outcome of any measurement that may be performed on
a system prepared by the associated preparation.} The point is that, if
one knows the state, one can predict probabilities for any
measurement that may be performed.  It is not entirely
clear that one will be able to ascribe states to preparations. The first
axiom, to be introduced later, will make this possible by assuming that
the same probability is obtained under the same circumstances.  If we
can ascribe a state it is clear from the definition above that one way
of describing the state is by that
mathematical object which simply lists all the probabilities for every
outcome of every conceivable measurement that could possibly be made on
the system.  This would be a very long list. Since most
physical theories have some structure, it is likely that this would be
too much information.  We can imagine that a set of $K$ appropriately chosen
probability measurements will be just sufficient and necessary to
determine the state (so $K$ is the smallest number of probabilities
required to specify the state). We will call these the
{\it fiducial measurements}.
We can list just the probabilities corresponding to these fiducial
measurements in the form of a column vector. Thus, the state can be written
\begin{equation}
  {\bf p} = \left(
  \begin{matrix} p_1 \\ p_2 \\ p_3 \\ \vdots \\ p_K \end{matrix}
  \right).
\end{equation}
We will call the integer $K$ the {\it number of degrees of freedom}. This
number plays an important role in this work.

The allowed states ${\bf p}$ will belong to some set $S$.  We expect
that there will exist sets of states which can be distinguished from each other
in this set by a single shot measurement.  Consider one such set. If
Alice picks a state from this set and sends it to Bob
then Bob can set up a measurement
apparatus such that each state gives rise to a disjoint set of outcomes.
By knowing which outcomes are associated with which state, Bob can tell
Alice which state she sent.  Let the maximum number of states in any
such set be called $N$.  We will call $N$ the {\it dimension} (because in
quantum theory it corresponds to the dimension of the Hilbert space).

Associated with any particular type of system will be the two integers
$K$ and $N$.  It turns out that in classical probability theory we have
$K=N$ and in quantum theory we have $K=N^2$.  We will explain why this
is the case later.

\begin{figure*}[t]
\resizebox{\textwidth}{!}
{\includegraphics{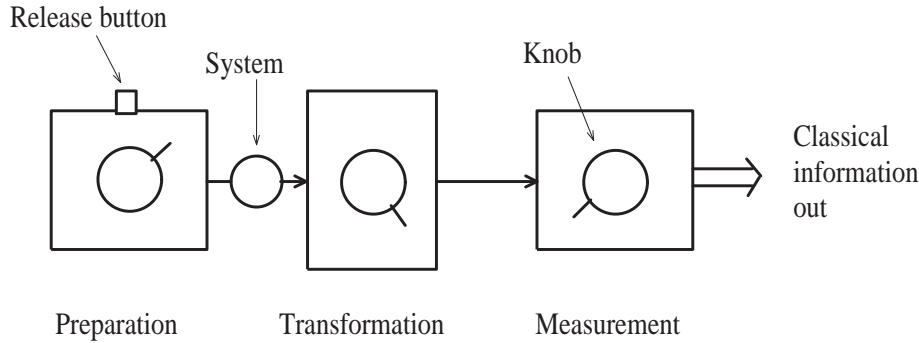}}
\caption{\small The situation considered consists of a preparation device with
a knob for varying the state of the system produced and a release button
for releasing the system, a transformation device for transforming the
state (and a knob to vary this transformation), and a measuring
apparatus for measuring the state (with a knob to vary what is measured)
which outputs classical information.}
\end{figure*}

First, let us describe the type of scenario we wish to consider. This is
shown in Fig.~1.  We have three types of apparatus. The preparation
apparatus prepares systems in some state.  It has a knob on it for
varying the type of state prepared. It also has a release button, whose
role will be described shortly.  The system then
passes through a transformation apparatus. This has a knob on
it which varies the transformation effected. Unless otherwise stated, we
will assume that the transformation device is set to leave the state
unchanged (i.e. effect the identity transformation).  Finally the system
impinges onto a measurement apparatus.  This has a knob on it to vary
the measurement being performed.  It also has some classical information
coming out. Either we obtain a non-null outcome, labeled $l=1$ to $L$, or we
obtain a {\it null outcome}.  We require that if the release button is
pressed on the preparation apparatus (and assuming that the
transformation is set to the identity) then we will certainly obtain a
non-null outcome. On the other hand, if the release button is not
pressed then we will certainly obtain a null outcome.  To illustrate
this we could think of an array of detectors labeled $l=1$ to $L$. If
none of the detectors click then we can say this is a null result.
Since we allow null outcomes we need not assume that states are
normalized.

All quantities are reducible to measurements of probability.  For
example, any measurement of an expectation value is really a probability
weighted sum.  Therefore, we need only consider measurements of probability.
Henceforth, when we refer to a ``measurement'' or a ``probability
measurement'' we mean specifically a measurement of the probability
that the outcome belongs to some non-null subset of outcomes with a
given knob setting on the measurement apparatus.

If we never press the release button then all the fiducial probability
measurements will be equal to zero (so the state will be represented by
a column vector with $K$ zero's). We will call this state the null state.

It is normal in probability theory to talk about pure states and mixed
states.  A mixed state is any state which can be simulated by a mixture
of two distinct states. Thus, we prepare randomly either state $A$ or
state $B$ with probabilities $\lambda$ and $1-\lambda$ where
$0<\lambda<1$.  Pure states are defined to be those states
(except the null state) which are not mixed states.  Pure states will
turn out to be extremal states in the set of allowed states (this set
being convex).

We will now describe classical probability theory and then quantum
theory.  We will find that it is possible to give the two theories a
very similar mathematical structure.  This will help us to appreciate the
similarities and differences between the two theories.

\section{Classical probability theory}

Consider a ball that can be in one of $N$ boxes (or be missing).  The
state is fully determined by specifying the probabilities, $p_n$, for
finding the ball in each box.  This information can, as in the previous
section, be written
\begin{equation}
  {\bf p} = \left(
  \begin{matrix} p_1 \\ p_2 \\ p_3 \\ \vdots \\ p_N \end{matrix}
  \right).
\end{equation}
Since the ball may be missing, the sum of the probabilities in this
vector must be less than or equal to one.  There are $N$ entries in
${\bf p}$. Hence, $K=N$.  There are some interesting special cases.  The
states
\begin{equation}
{\bf p}_1= \left(
\begin{matrix} 1 \\ 0 \\ 0 \\ \vdots \\ 0 \end{matrix} \right) \qquad
{\bf p}_2= \left(
\begin{matrix} 0 \\ 1 \\ 0 \\ \vdots \\ 0 \end{matrix} \right) \qquad
{\bf p}_3= \left(
\begin{matrix} 0 \\ 0 \\ 1 \\ \vdots \\ 0 \end{matrix} \right) \qquad
{\rm etc.}
\end{equation}
represent the case where the ball is definitely in one of the boxes.
These states cannot be simulated by mixtures of other states and hence
are pure states for this system.
The state
\begin{equation}
{\bf p}_{\rm null}= {\bf 0} =
\left( \begin{matrix} 0 \\ 0 \\ 0 \\ \vdots \\ 0 \end{matrix} \right)
\end{equation}
represents the case where the ball is missing.
These $N+1$ states are extremal in the space of allowed states.
Since we are casting classical
probability theory and quantum theory in similar mathematical forms,
let us consider how we can represent measurements in the classical case.
One measurement we could make is to look and see if the ball is in box
1.  The probability of finding the ball in box 1 is $p_1$. We can write
this as
\begin{equation}
p_1=
\left(   \begin{matrix} 1 \\ 0 \\ 0 \\ \vdots \\ 0 \end{matrix} \right)
\cdot
\left(   \begin{matrix} p_1 \\ p_2 \\ p_3 \\ \vdots \\ p_N \end{matrix} \right)
={\bf r}_1\cdot {\bf p}.
\end{equation}
Hence, we can identify the vector ${\bf r}_1$, defined as
\begin{equation}
{\bf r}_1= \left(
\begin{matrix} 1 \\ 0 \\ 0 \\ \vdots \\ 0 \end{matrix} \right),
\end{equation}
with the measurement where we look to see if the ball is in box 1. We
can write down similar vectors for the other boxes.  However, we could
perform more complicated measurements.  For example, we could toss a
$\lambda$ biased coin and look in box 1 if it came up heads and in box 2
if it came up tails.  In this case the measurement being performed would
be represented by the vector
\begin{equation}
{\bf r}=\lambda{\bf r}_1+(1-\lambda){\bf r}_2
\end{equation}
since then ${\bf r}\cdot{\bf p}=\lambda p_1 + (1-\lambda)p_2$.
In general it can be shown that the probability associated with any
measurement is given by
\begin{equation}
{\rm prob}_{\rm meas} = {\bf r}\cdot{\bf p}
\end{equation}
where ${\bf r}$ is associated with the measurement and ${\bf p}$ is
associated with the state.

Consider a classical
bit. This is a system with $N=2$.  In this case the extremal states
are
\begin{equation}
{\bf p}_1= \left( \begin{matrix} 1 \\ 0 \end{matrix} \right) \qquad
{\bf p}_2= \left( \begin{matrix} 0 \\ 1 \end{matrix} \right) \qquad
{\bf p}_{\rm null}= \left( \begin{matrix} 0 \\ 0 \end{matrix} \right).
\end{equation}
The set of allowed states $S_{\rm classical}$ are given by the convex
hull of these extremal states as shown
in Fig.~2a. Note that the normalized states (for which $p_1+p_2=1$)
lie on the hypotenuse.  Note also that the pure states form a discrete
set.  There is no continuous path from one pure state to another which
goes through the pure states.

We see that classical probability theory is characterized by $K=N$, by the
set $S_{\rm classical}$ of allowed states ${\bf p}$ and the set
$R_{\rm classical}$ of allowed measurements ${\bf r}$,
and by the formula ${\rm prob}_{\rm meas}={\bf r}\cdot{\bf p}$.
\begin{figure*}[t]
\resizebox{\textwidth}{!}
{\includegraphics{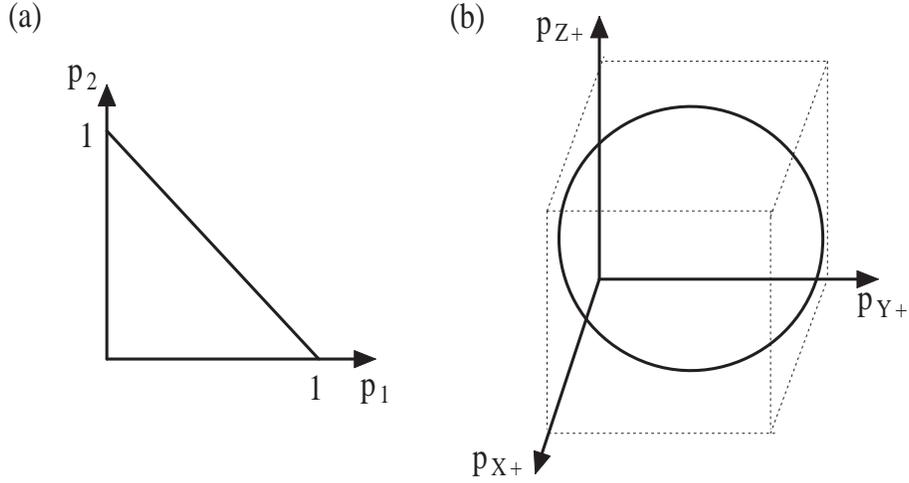}}
\caption{\small (a) Allowed states for classical bit are inside triangle.
States on the hypotenuse are normalized.  (b) Normalized states for a
qubit are in the ball inside the unit cube as shown.}
\end{figure*}

\section{Quantum theory}\label{quantumtheory}

Let us begin describing the quantum case by discussing an example.
Consider a spin half particle (an example of a qubit).  Its state is
represented by a density matrix $\rho$ which can be written
\begin{equation}
\rho= \left( \begin{matrix} p_{z+} & a \\ a^* & p_{z-} \end{matrix}
\right)
\end{equation}
where
\begin{equation}
a=p_{x+} - p_{y+} - {1-i\over 2}(p_{z+}+p_{z-}).
\end{equation}
Here, $p_{z+}$ is the probability the particle has
spin up along the $+z$ direction and the other probabilities are defined
similarly.  This means that rather than
representing the state by $\rho$, we can represent it by
\begin{equation}
{\bf p} = \left( \begin{matrix} p_{z+} \\ p_{z-} \\ p_{x+} \\ p_{y+}
\end{matrix} \right).
\end{equation}
This mathematical object contains the same information as $\rho$.
Hence, for $N=2$ we have $K=4$.  The set of allowed states can be
calculated from the condition that $\rho$ is positive.  Since there are
four parameters it is not easy to visualize the shape of this set.
However, if we impose normalization ($p_{z+}+p_{z-}=1$), thus eliminating
one variable, then we can picture the allowed set of states in three
dimensions.  We find that the allowed states are inside that ball which sits
just in the unit cube in the first octant of the variables
$p_{x+}, p_{y+}, p_{z+}$ as shown in Fig.~2b.  This is basically the
Block sphere in a different coordinate set.  All the points on the
surface of the ball represent pure states (since they are extremal).
Hence, unlike in the classical case, the pure states form a continuous
set.  This will be the key difference between the two theories.

The density matrix for $N=2$ is specified by 4 real parameters and this
is why we need four probabilities.  In general, the density matrix for a
system of dimension $N$ is specified by $N^2$ real parameters (since we have
$N$ real numbers along the diagonal and $N(N-1)/2$ complex numbers above
the diagonal).  Not surprisingly then, we can show that we
need $N^2$ probabilities to describe a general state:
\begin{equation}
  {\bf p} = \left(
  \begin{matrix} p_1 \\ p_2 \\ p_3 \\ \vdots \\ p_{N^2} \end{matrix}
  \right).
\end{equation}
Hence, $K=N^2$.  Various authors have noticed that the state can be
represented by probabilities \cite{wootters,prug,busch,stefan}.

Associated with each probability measurement in quantum theory is a
positive operator $A$.  The probability for that measurement is given by
the trace formula:
\begin{equation}
{\rm prob}_{\rm meas} = {\rm tr}(A\rho)
\end{equation}
Now, since $\rho$ is linear in the probabilities $p_k$ for $k=1$ to
$N^2$, it follows that we can write
\begin{equation}
{\rm prob}_{\rm meas} = {\bf r}\cdot {\bf p}.
\end{equation}
The vector ${\bf r}$ can be determined from $A$. It describes the
measurement.

Quantum theory is characterized by $K=N^2$,
by the set $S_{\rm quantum}$ of allowed states ${\bf p}$ and the set
$R_{\rm quantum}$ of allowed measurements ${\bf r}$,
and by the formula ${\rm prob}_{\rm meas}={\bf r}\cdot{\bf p}$.

It is also interesting to think about the effect of the transformation
device on the state.  In quantum theory, transformations are described by
unitary transformations or, in the case of open systems, by
superoperators.  When acting on ${\bf p}$, it can be shown that such
transformations can be written
\begin{equation}
{\bf p} \longrightarrow Z{\bf p}
\end{equation}
where $Z$ is a $K\times K$ real matrix.  (A similar statement holds for
transformations in the case of classical probability theory.)
Allowed transformations belong to some set $Z\in \Gamma_{\rm
quantum}$.

\section{The axioms}

We will soon state the five axioms.  But first let us point out a number
of features of classical probability theory and quantum theory.
Both theories are probability theories.  We can only build a useful
theory of probability if the world is such that the same probability is
obtained under the same circumstances.  Axiom 1 imposes this condition.
The remaining axioms impose restrictions on the structure of the
probability theory we derive.  To motivate Axiom 2 consider
the situation where a ball can be in one of five boxes. Then
$N=5$.  However, if the state is constrained so that the ball is never
found in the last two boxes then the system will behave like one with
$N=3$.  Similarly, if the state of a quantum system is constrained to a lower
dimensional subspace of the Hilbert space then it will behave like a
system of the dimension of the subspace. We will say, in general, that a
state is constrained to an $M$ dimensional subspace if, with the
measurement apparatus set to distinguish a set of $N$ distinguishable
states, the only outcomes observed (apart from the null outcome)
are those associated with a subset of
$M$ of these distinguishable states.  In both classical and quantum
theory the system will behave like one of dimension $M$ in such cases.
To motivate the third axiom
consider a composite system consisting of systems $A$ and $B$.  In both
classical and quantum theory we have that $N=N_AN_B$ and $K=K_AK_B$.
One set of functions $K=K(N)$ which satisfy these properties are $K=N^r$
where $r$ is a positive integer.  In fact, it will turn out from the
axioms that $K(N)$ must be of this form.  The simplest case is $K=N$ (with
$r=1$).  This is consistent with classical probability theory.  However,
the fourth axiom will imply that there exists a continuous set of pure
states.  This rules out $K=N$.  The next simplest case is $K=N^2$.  This
corresponds to quantum theory.  The role of Axiom 5 will be to take the
simplest case consistent with the constraints imposed by the
axioms (namely $K=N^2$).

The five axioms for quantum theory are:
\begin{description}
\item[Axiom 1] {\it Probabilities}.  Relative frequencies (measured by
taking the proportion of times a particular outcome is observed)
tend to the same value (which we call the probability) for any case
where a given measurement is performed on a ensemble of $n$ systems
prepared by some given preparation in the limit as $n$ becomes infinite.
\item[Axiom 2] {\it Subspaces}. There exist systems for which
$N=1,2,\cdots$, and, furthermore, all systems of dimension $N$, or
systems of higher
dimension but where the state is constrained to an $N$ dimensional
subspace, have the same properties.
\item[Axiom 3]  {\it Composite systems}. A composite system consisting of
subsystems $A$ and $B$ satisfies $N=N_AN_B$ and $K=K_AK_B$.
\item[Axiom 4] {\it Continuity}. There exists a continuous reversible
transformation on a system between any two pure states of that
system for systems of any dimension $N$.
\item[Axiom 5] {\it Simplicity}. For each given $N$, $K$ takes the
minimum value consistent with the other axioms.
\end{description}
The axioms are written in a slightly different (though obviously
equivalent) form to those given in \cite{Hardy1}.
If the word ``continuous'' is dropped from Axiom 4 then, because of the
simplicity axiom, we obtain classical probability theory instead of
quantum theory.  It is rather striking that the difference between
classical probability theory and quantum theory is just one word.

A few comments on these axioms are appropriate here.  We can think of
any probability theory as a structure.  This structure, however, has no
physical meaning unless we have a way of relating it to the real world.
The first axiom deals with this aspect.  It states that probabilities,
defined as limiting relative frequencies, are the same each time they
are measured.  There are various different interpretations of
probability.  Axiom 1, as stated, favours the frequency approach.
However, one could recast this axiom in keeping with other
interpretations such as the Bayesian approach \cite{schack}.
In this paper we are
primarily concerned with the structure of quantum theory and so will
not try to be sophisticated with regard to the interpretation of
probability theory.  However, these are important matters which
deserve further attention.

By a ``continuous transformation'' we mean one that can be built up of many
transformations which are themselves only infinitesimally different from
the identity transformation.
The motivation for the continuity axiom is simply that we would like
physics to be continuous.  There is no way, in finite dimensional
classical probability theory, of going in a continuous way from one pure
state to another.  It is classical probability theory that has the
``jumps''.

The motivation for $N=N_AN_B$ is fairly clear.  For example, if we have
two dice then $N_A=N_B=6$ and $N=36$.  However, the motivation for
$K=K_AK_B$ is not so clear.  It follows from two intuitions.
Intuition A: Pure states represent definite states.  This motivates
Assumption $\alpha$: If one of the two subsystems is in a pure
state then any joint
probabilities factorize (since a system in a definite state should not
be correlated with any other system).  From this we can show that the
number of degrees of freedom associated with the separable states (those
states that can be regarded as a mixture of states whose joint probabilities
factorize) is
\begin{equation}
K_{\rm separable} = K_A K_B.
\end{equation}
Intuition B: There should not be more entanglement than necessary.
This motivates Assumption $\beta$: $K=K_{\rm separable}$.
Hence, $K=K_AK_B$ follows.

The simplicity axiom has a slightly awkward status.  It is perhaps
better regarded as a meta-axiom (applied to a set of axioms).  As a
guiding principle in physics, simplicity is perfectly valid. However, it
would more satisfactory to either show that theories with $K=N^r$ for
$r>2$ do not exist or that they can be ruled out by adding some
additional reasonable axiom.  On the other hand, if such theories do
exist, then it would be very interesting to actually construct them and
investigate their properties.  It may turn out that they have even
better information processing capacity than quantum theory.
Furthermore, there may be a downward compatibility. Thus, classical
probability theory can be embedded in quantum theory (by only taking
orthogonal states).  It may be that quantum theory can be embedded in a
higher power theory.  If this turned out to be the case then such a
theory may be consistent with all the empirical data collected to date
and could therefore be a true theory of the world.

\section{Derivation of quantum theory from the axioms}

The proof that these axioms give quantum theory is rather complicated
and so we will content ourselves here with simply indicating how the
various steps of the proof work. The reader is referred to \cite{Hardy1}
for details of the proofs.

\subsection{Linearity}

It follows from Axiom 1 that measured probabilities do not depend on the
particular ensemble being used. Thus, we can associate a state, ${\bf
p}$, with a
preparation as discussed in Section \ref{basicnotions}.
The probability associated
with a general measurement will be given by some function of the state:
\begin{equation}
{\rm prob}_{\rm meas} = f({\bf p}).
\end{equation}
This function will, in general, be different for each measurement.
Let ${\bf p}_C$ be the mixed state prepared when state ${\bf p}_A$ is
prepared with probability $\lambda$ and state ${\bf p}_B$ is prepared
with probability $1-\lambda$.  Then we have
\begin{equation}\label{convexf}
f({\bf p}_C) = \lambda f({\bf p}_A)
+(1-\lambda) f({\bf p}_B).
\end{equation}
We can apply this equation to the fiducial measurements
themselves. This gives
\begin{equation}\label{convexp}
{\bf p}_C = \lambda {\bf p}_A+(1-\lambda){\bf p}_B
\end{equation}
since this equation is true for each component by (\ref{convexf}).
Hence,
\begin{equation}\label{convexpf}
f(\lambda {\bf p}_A+(1-\lambda){\bf p}_B)
= \lambda f({\bf p}_A)+(1-\lambda) f({\bf p}_B)
\end{equation}
This can be used to prove that the function $f$ is linear in ${\bf p}$.
Hence, we can write
\begin{equation}
{\rm prob}_{\rm meas}= {\bf r}\cdot {\bf p}
\end{equation}
where ${\bf r}$ is a vector associated with the measurement.
It follows from (\ref{convexp}) that the set of allowed states $S$ must
be convex.  The extremal states (except the null state) are the pure
states.  They cannot be written as a mixture of any other states.

\subsection{Proof that $K=N^r$}

Axiom 2 says that
any system of dimension $N$ has the same properties. This implies that
$K=K(N)$.  From Axiom 3 we can write
\begin{equation}\label{compmult}
K(N_AN_B)=K(N_A)K(N_B).
\end{equation}
Such functions are known in number theory as {\it completely
multiplicative}.  It follows from the subspace axiom that
\begin{equation}\label{kgtk}
K(N+1)>K(N).
\end{equation}
From (\ref{compmult}, \ref{kgtk}) it can be proven that
\begin{equation}
K=N^{\alpha}
\end{equation}
where $\alpha>0$.  Since $K$ is an integer we must have $K=N^r$ where
$r=1,2,\cdots$. Wootters, employing related reasoning, has also
come to the equation $K=N^r$ as a possible relationship between $K$ and
$N$ \cite{wootters}.

The simplicity axiom requires that we take the smallest value of $r$
consistent with axioms 1 to 4.  If we drop the word ``continuous'' from
Axiom 4 then this gives $K=N$ and it can be shown that
we obtain classical probability
theory.  However, it can be shown that the $K=N$ case cannot give rise
to a continuous set of pure states. Hence, if Axiom 4 is left as it
is then we must, by the simplicity axiom, have $K=N^2$.

\subsection{Qubits}

We can consider the case where $N=2$.  Since $K=N^2$ we then
have $K=4$.  One of these degrees of freedom is associated with
normalization.  If we consider normalized states we have only three
degrees of freedom.  Axiom 4 requires that there exist continuous
reversible transformations between any two pure states.  These
reversible transformations will form a group. It can be shown that they
generate a set of pure states which are on the surface of a ball
corresponding exactly to the quantum case (discussed in Section
\ref{quantumtheory}).

\subsection{General $N$}

Having obtained quantum theory for the special case $N=2$ we can use
this in conjunction with the subspace axiom to recover quantum theory
for general $N$.  We do this by considering two dimensional subspaces.
We require that, if the state is restricted to any given two dimensional
subspace, then it behaves like a qubit.   With this constraint we can
obtain the trace formula for predicting probabilities and the
constraints that ${\bf r}$ and ${\bf p}$ correspond to the positive
operators $A$ and $\rho$ respectively.

\subsection{Transformations}

It can be shown from linearity that transformations are of the form
${\bf p}\rightarrow Z{\bf p}$ where $Z\in\Gamma$ is a $K\times K$ real
matrix.
By considering composite systems we can find the most general class of
transformations consistent with the axioms.  These turn out to
correspond to the completely positive linear trace non-increasing maps
of standard quantum theory \cite{krauss,nielsenchuang}.

\subsection{State update rule}

One of the more mysterious features of quantum theory is the state update
rule. If the system emerges from the measurement apparatus its state
will, in general, have changed.  In text books the von Neumann
projection principle is usually given. However, this is by no means the
most general state change that can happen after a measurement.  In
general, we expect each outcome $l$ to be associated with a particular
transformation $Z_l\in\Gamma$ of the state. The normalization associated
with the state after a particular measurement result will be consistent
with the probability for that outcome. These transformations can all be
taken together. Hence we require $\sum_l Z_l \in \Gamma$.  These
constraints are sufficient to give the most general state update rule of
quantum theory.  It is  interesting to note that exactly the same
constraints apply in
classical probability theory.  Thus, the strangeness associated with the
state update rule in quantum theory is not so much due the way in which
the state is updated as it is due to the nature of the sets $S$ and $R$
in quantum theory.

\section{Discussion}

The basic property from which quantum theory follow is that there should
be continuous transformations between pure states.  In classical
probability theory for discrete systems it is necessary to
jump between pure states.  We
might ask what would have happened had a nineteenth century physicist
complained about ``dammed classical jumps''.  It is possible that he
would have gone on to develop quantum theory.  There is a sense in
which quantum theory is more reasonable than classical theory exactly
because there do exist these continuous transformations.

There are various reasons for developing reasonable axioms.  Firstly,
physics is primarily about explanation and we can be said to have
explained quantum theory more deeply if we give reasonable axioms.
Secondly, by having a deeper understanding of the origin of quantum
theory we are more likely to be able to extend or adapt the theory to
new domains of applicability (such as quantum gravity).  Thirdly, the
fact that we put quantum theory and classical probability theory on such
a similar footing may point the way to a deeper appreciation of
the relationship between classical and quantum information.
And finally, these new axioms may shed some light on the interpretation
of quantum theory.

\vspace{6mm}

\noindent{\Large\bf Acknowledgements}

\vspace{6mm}

This work is funded by a Royal Society University Research Fellowship.

\end{document}